\documentstyle[12pt]{article}
\def\eqlab#1{\label{eqn:#1}}
\def\eqref#1{Eq.(\ref{eqn:#1})}

\def\PR#1#2#3{Phys. Rev. {\bf #1} (#3) #2 }
\def\PRL#1#2#3{Phys. Rev. Lett. {\bf #1} (#3) #2 }
\def\PL#1#2#3{Phys. Lett. {\bf #1} (#3) #2 }

\def\PTP#1#2#3{Prog. Theor. Phys. {\bf #1} (#3)#2 }

\begin{document}
\renewcommand{\thefootnote}{\fnsymbol{footnote}}
\renewcommand{\theenumi}{(\roman{enumi})}
%\renewcommand{\theequation}{\thesection.\arabic{equation}}
% ------------------------------------------------------------------------ %
\title{Is $U_{e3}$ really related to the solar neutrino solutions?}
\author{
{N. Haba$^{1,2}$}\thanks{haba@eken.phys.nagoya-u.ac.jp}
{, and Tomoharu Suzuki$^2$}\thanks{tomoharu@eken.phys.nagoya-u.ac.jp}
\\
\\
\\
{\small \it $^1$Faculty of Engineering, Mie University,}
{\small \it Tsu Mie 514-8507, Japan}\\
{\small \it $^2$Department of Physics, Nagoya University,}
{\small \it Nagoya, 464-8602, Japan}}
%\date{May 8, 2000}
\date{}
\maketitle
% ------------------------------------------------------------------------ %
%\vspace{-12.5cm}
\vspace{-10.5cm}
\begin{flushright}
hep-ph/0006281\\
DPNU-00-22\\
%KEK-TH-691\\
\end{flushright}
\vspace{10.5cm}
\vspace{-2.5cm}
\begin{center}
\end{center}
% ------------------------------------------------------------------------ %
\renewcommand{\thefootnote}{\fnsymbol{footnote}}
%\renewcommand{\thefootnote}{\arabic{footnote}}
% ======================================================================== %
%
% abstract
%
% ------------------------------------------------------------------------ %
\begin{abstract}

It has been said that the measurements of $U_{e3}$ in the 
 lepton flavor mixing matrix would help discriminate between 
 the possible solar neutrino solutions under the 
 natural conditions with the neutrino mass hierarchies of 
 $m_1 \ll m_2 \ll m_3$ and $m_1 \sim m_2 \gg m_3$, where
 $m_i$ is the $i$-th generation neutrino absolute mass. 
However, it is not true, and 
 the relation between $\sin^2 2 \theta_{12}$ and $U_{e3}$ 
 obtained by Akhmedov, Branco, and Rebelo 
 is trivial in actual. 
We show in this paper that 
 the value of $U_{e3}$ cannot predict  
 the solar neutrino solutions
 without 
 one additional nontrivial 
 condition. 
%The value of $U_{e3}$ is the physical 
% quantity which can be measured 
% by the experiments independently of 
% the solar neutrino solutions. 
%In order for $U_{e3}$ to 
% predict the solar neutrino solution, 
% one additional nontrivial 
% theoretical information must be needed. 

\end{abstract}

\newpage
% ======================================================================== %
%\section{Introduction}
% ------------------------------------------------------------------------ %

Recent neutrino oscillation experiments 
 suggest the strong evidences of
 tiny neutrino masses and 
 lepton flavor mixings\cite{solar4,Atm4,SK4,CHOOZ}. 
Studies of the lepton flavor mixing matrix,
 which is so-called Maki-Nakagawa-Sakata (MNS) matrix\cite{MNS},
 will give us important cues of the physics
 beyond the standard model. 
The mixing angle between the second and the third 
 generations is expected to be almost maximal\cite{SK4}, and 
 the large mixing between the first and the second 
 generations is also favored\cite{nu2000} as the large angle 
 MSW (MSW-L) solution\cite{MSW}. 
On the other hand, 
 the mixing between the first and the third 
 generations, which corresponds to $U_{e3}$ in the 
 MNS matrix, is small 
 as the present upper bound of CHOOZ experiments 
 show $U_{e3}< 0.16$\cite{CHOOZ}. 
It is very interesting if value of $U_{e3}$ is 
 related to the solar neutrino solutions. 
In Ref.\cite{ABR}, Akhmedov, Branco, and Rebelo 
 said that the measurements of $U_{e3}$ 
 would help discriminate between 
 the possible solar neutrino solutions under the 
 natural conditions with the neutrino mass hierarchies of 
 $m_1 \ll m_2 \ll m_3$ and $m_1 \sim m_2 \gg m_3$, where
 $m_i$ is the $i$-th generation neutrino absolute mass. 
However, it is not true, and 
 the relation between $\sin^2 2 \theta_{12}$ and $U_{e3}$ 
 obtained in Ref.\cite{ABR} is trivial in actual. 
We will show in this paper that 
 the value of $U_{e3}$ cannot predict 
 the solar neutrino solutions 
 without one additional nontrivial 
 condition. 
This is because we know only four parameters, 
 $\sin^2 2 \theta_{12}, \sin^2 2 \theta_{23}, 
 \Delta m^2_{sol}$, and 
 $\Delta m^2_{ATM}$, from experiments, 
 although five parameters must be needed in order to
 obtain the MNS matrix, and its element $U_{e3}$.

\vspace{5mm}
% ======================================================================== %
%\section{Neutrino mass matrix}
% ------------------------------------------------------------------------ %

Let us start our discussions with each type of 
 neutrino mass hierarchy. 
Neutrino mass spectra can be classified in 
 three types\cite{type} as, Type A:  $m_1  \ll m_2 \ll m_3$, 
 Type B: $m_1 \sim m_2 \gg m_3$, and 
 Type C: $m_1 \sim m_2 \sim m_3$. 
It is expected that $\Delta m_{ATM}^2 \simeq |m_3^2 - m_2^2 |$ 
 and $\Delta m_{sol}^2 \simeq |m_2^2 - m_1^2 |$. 
By using $\theta_{23} = \pi /4$ and 
 $U_{e3}= \epsilon ( \ll 1)$ 
 according to the data of the 
 Super-Kamiokande and the CHOOZ experiments, 
 respectively, 
 the MNS matrix  
 is given by \cite{Ak} 
\begin{equation}
U = 
\left(\begin{array}{ccc}
c & s & \epsilon \\
-{1 \over \sqrt{2}}(s+c \epsilon ) & 
 {1 \over \sqrt{2}}(c-s \epsilon ) & {1 \over \sqrt{2}} \\
 {1 \over \sqrt{2}}(s-c \epsilon ) 
 & -{1 \over \sqrt{2}}(c+s \epsilon ) & {1 \over \sqrt{2}}
\end{array}\right)
 \,,
\eqlab{charged-lepton}
\end{equation}
where $c \equiv \cos \theta_{12}$ and $s \equiv \sin \theta_{12}$. 
The Majorana 
 mass matrix of neutrino in the diagonal 
 base of the charged lepton mass matrix
 is given by \cite{Ak}
\begin{eqnarray}
\label{M_nu}
M_{\nu}
& = & U \; diag.(m_1, m_2, m_3) \; U^T \; , \\
& = & \left(
\begin{array}{ccc}
\mu  & {1 \over \sqrt{2}}[ \epsilon (m_3- \mu)+m_{-}cs] & 
       {1 \over \sqrt{2}}[ \epsilon (m_3- \mu)-m_{-}cs]   \\
 {1 \over \sqrt{2}}[ \epsilon (m_3- \mu)+m_{-}cs]  & 
 {1 \over 2}(m_3 + \mu' -2 m_{-}cs \epsilon )  & 
 {1 \over 2}(m_3 - \mu') \\
 {1 \over \sqrt{2}}[ \epsilon (m_3- \mu)-m_{-}cs] &
 {1 \over 2}(m_3 - \mu') & 
{1 \over 2}(m_3 + \mu' -2 m_{-}cs \epsilon )  
\end{array}
\right), 
\end{eqnarray}
\vspace{-0.3cm}
where
\begin{equation}
\mu \equiv m_1 c^2 + m_2 s^2, \;\;\;\;\;\;\;
\mu' \equiv m_1 s^2 + m_2 c^2, \;\;\;\;\;\;\;
m_{-} \equiv m_2 - m_1. 
\end{equation}

\vspace{5mm}
% ======================================================================== %
%\section{Type A ($m_1 \ll m_2 \ll m_3$)}
% ------------------------------------------------------------------------ %

In Type A with 
 the mass hierarchy of $m_1 \ll m_2 \ll m_3$,
 the neutrino mass matrix of Eq.(3) 
 is written by 
\begin{eqnarray}
\label{Mnu}
M_{\nu}
& = & m_0
\left(
\begin{array}{ccc}
 \kappa          & \alpha + \beta & \alpha - \beta  \\ 
 \alpha + \beta  & 1 + \delta - \delta' & 1 - \delta \\
 \alpha - \beta  & 1 - \delta & 1 + \delta + \delta'
\end{array}
\right), 
\end{eqnarray}
where we just normalize Eq.(3) by 
 $m_3$ as 
%
%\begin{equation}
$m_0 \equiv {m_3 / 2}$, 
 $\kappa \equiv {2 \mu / m_3}$, 
 $\alpha \equiv \sqrt{2} \epsilon (
     1 - {\mu / m_3})$, 
 $\beta \equiv {\sqrt{2} m_{-}cs / m_3}$, 
 $\delta \equiv {\mu' / m_3}$,
and
$
 \delta' \equiv ({2 m_{-}cs / m_3}) \epsilon .
$
%\end{equation}
% 
The values of $m_0, \kappa, \beta$, and $\delta$ 
 are determined by the atmospheric and the solar
 neutrino solutions. 
Only $\alpha$ and $\delta'$ are
 unknown parameters, since they have the free parameter 
 $\epsilon$. 
Equation (\ref{Mnu}) induces 
 the mixing angles of 
\begin{eqnarray}
& & \tan 2 \theta_{12} = 
   {\sqrt{2} \beta \over \delta 
      - {\kappa \over 2}}, \;\;\;\;\;\;\;\;\;\;
\sin \theta_{13} = 
   {\alpha \over \sqrt{2} \left( 1- {\kappa \over 2}\right)}.  
\end{eqnarray}
By using the approximations of 
 $\Delta m^2_{ATM}\simeq m_3^2$ and 
 $\Delta m^2_{sol}\simeq m_2^2$, 
 they become 
\begin{eqnarray}
\label{8}
& & \tan 2 \theta_{12} \simeq 
   {\sqrt{2} \beta \over 
    \cos 2 \theta_{12} \sqrt{\Delta m^2_{sol} 
     \over \Delta m^2_{ATM}}}, \;\;\;\;\;\;\;\;\;
\sin \theta_{13} \simeq 
   {\alpha \over \sqrt{2} \left( 1- \sin^2 \theta_{12}
   \sqrt{\Delta m^2_{sol} \over \Delta m^2_{ATM}} \right)}.  
\end{eqnarray}
Here we must notice that the value of $\beta$ is
 determined by 
 the atmospheric and
 the solar neutrino solutions.
Only $\epsilon = \sin \theta_{13}$ is 
 the free parameter with 
 $U_{e3}\;(= \; \sin \theta_{13}\;) 
 \; <0.16$\cite{CHOOZ}, which 
 makes the value of $\alpha$ be also
 free parameter. 
If $O( \alpha ) \simeq O( \beta )$, which dose not 
 have physical meaning, Eqs.(\ref{8}) induce
\begin{equation}
\label{sintan1}
U_{e3} 
   \simeq {1 \over \sqrt{2}}\;
     \sin 2 \theta_{12}\; \sqrt{\Delta m^2_{sol} 
     \over \Delta m^2_{ATM}} .
\end{equation}
The right-hand side of 
 this equation\footnote{
Equation (\ref{sintan1}) is not the same as 
 the result in Ref.\cite{ABR} 
$
\sin \theta_{13} \simeq {1 \over 2}\;
     {\tan 2 \theta_{12} \over 
      (1+ \tan^2 2 \theta_{12})^{1/4}}
     \left({\Delta m^2_{sol} 
     \over \Delta m^2_{ATM}}\right)^{1/2} ,
$
which can not apply to the large angle solutions. 
Our result of Eq.(\ref{sintan1}) can apply not only to 
 the small angle solution but also to 
 the large angle solutions. 
}
 gives the following values of $U_{e3}$ 
 corresponding to 
%the atmospheric and
 the solar neutrino solutions as 
\begin{eqnarray}
\label{10}
U_{e3} & \sim & 
   10^{-1.5}\;\;\; {\rm (MSW-L)}, \;\;\;\;\; 
   10^{-3.5}\;\;\; {\rm (VO)}, \;\;\;\;\; 
   10^{-3}\;\;\; {\rm (MSW-S)}.  
\end{eqnarray}
These results are the same as those of Ref.\cite{ABR}. 
It seems that the measurements of $U_{e3}$ 
 can predict the solar neutrino solutions from Eq.(\ref{10}).  
However, we must notice that 
 the relation of Eq.(\ref{10}) is satisfied just only in 
 the case of $O( \alpha ) \simeq O( \beta )$. 
This is 
 the trivial condition, since 
 $\alpha$ is the free parameter which 
 has nothing to do with $\beta$ at all. 
In Ref.\cite{ABR}, 
 they have denoted $\varepsilon \equiv \alpha + \beta$ 
 and $\varepsilon' \equiv \alpha - \beta$, and 
 said that $\varepsilon + \varepsilon'$ and 
 $\varepsilon - \varepsilon'$ are 
 expected to be of the same order if there are no 
 accidental cancellations.
However, the condition $\varepsilon + \varepsilon' \simeq 
 \varepsilon - \varepsilon'$ means $\alpha \simeq \beta$, 
 which is not the natural condition 
 but just the trivial assumption. 
Any relations between $\alpha$ and $\beta$ 
 are considerable, and for example,  
 if we take $O( \alpha ) \ll O( \beta )$, 
 Eq.(\ref{sintan1}) becomes 
\begin{equation}
\label{sintan2}
\sin \theta_{13} \ll 
 {1 \over \sqrt{2}}\;
     \sin 2 \theta_{12}\sqrt{\Delta m^2_{sol} 
     \over \Delta m^2_{ATM}}. 
\end{equation}
We stress again that Eqs.(\ref{sintan1}) 
 and (\ref{sintan2}) are just the trivial relations 
 which are induced by the trivial assumptions of 
 $O( \alpha ) \simeq O( \beta )$ and 
 $O( \alpha ) \ll O( \beta )$, respectively.

\vspace{5mm}
% ======================================================================== %
%\section{Type B ($m_1 \sim m_2 \gg m_3$)}
% ------------------------------------------------------------------------ %

We can similarly analyze the case of Type B\footnote{
Type B has two patterns of (b1) and (b2) according to 
 the relative sign assignments of mass eigenvalues\cite{type}. 
The stability of the mixing angles 
 against the quantum corrections 
 strongly depends 
 on the relative assignments of mass eigenvalues\cite{HO1}. 
} 
 with the mass hierarchy of $m_1 \sim m_2 \gg m_3$. 
In the case of (b1), which is 
 $M_{\nu}^d \simeq diag.(m_1, m_2, 0)$ in the first order,  
 with the notation of $m_2 = m_1+d$ $(|d| \ll m_1)$ 
 the neutrino mass matrix $M_\nu$ is given by 
\begin{eqnarray}
M_\nu 
  &=&
  m_1
  \left(
  \begin{array}{ccc}
  1+\kappa & \alpha +\beta & \alpha -\beta \\
  \alpha +\beta & \frac{1}{2} +\delta +\gamma+\delta' 
                & -\frac{1}{2}+\delta-\gamma \\
  \alpha -\beta & -\frac{1}{2}+\delta-\gamma 
               & \frac{1}{2}+\delta +\gamma+\delta' 
  \end{array}
  \right) ,
\label{14}
\end{eqnarray}
\vspace{-0.2cm}
where 
%
%\begin{equation}
$
 \kappa \equiv {d s^2}/{m_1}$,  
 $\alpha \equiv -({\epsilon}/{\sqrt{2}})
  \left[
  1 + (m_3-ds^2)/{m_1}\right]$, 
 $\beta \equiv {dcs}/{\sqrt{2} m_1}$, 
 $\delta \equiv {m_3}/{2m_1}$, 
 $\gamma \equiv {d c^2}/{2 m_1}$, 
and
$
 \delta' \equiv ({d cs}/{m_1})\epsilon . 
$
%\end{equation}
%
The values of $\kappa, \beta, \delta$, and $\gamma$ 
 are determined by the atmospheric and the solar
 neutrino solutions. 
Only $\alpha$ and $\delta'$ are
 free parameters, since they contain $\epsilon$. 
We can easily obtain mixing angles 
\begin{eqnarray}
 \tan 2 \theta_{12}=\frac{\sqrt{2}\beta}
           {\gamma+\frac{\kappa}{2}}
      \; , \;\;\;\;\;\;\;\;
 \sin \theta_{13}=
       -\frac{\sqrt{2}\alpha}
            {1+\kappa-2\delta} \; ,
\end{eqnarray}
{}from Eq.(\ref{14}).
By using the approximations of
%
%\begin{eqnarray}
$
 \Delta m^2_{ATM} \simeq m_1^2, 
$
and
$
 \Delta m^2_{sol} 
       \simeq  4 m_1^2 
      ( \gamma+\frac{\kappa}{2})
        (1+\tan^2\theta_{12})^{1/2}, 
$
%\end{eqnarray}
%
we can obtain 
\begin{eqnarray}
 \alpha \simeq 
    -\frac{1}{\sqrt{2}} \sin \theta_{13} 
            \;,\;\;\;\;\;\;\;\;
    \beta \simeq  \frac{1}{4\sqrt{2}} 
                \sin 2\theta_{12} 
      \left(
        \frac{\Delta m^2_{sol}}{\Delta m^2_{ATM}}
      \right) .
\label{17}
\end{eqnarray}
\vspace{-0.2cm}
Here we must notice that the value of $\beta$ is
 determined by 
 the atmospheric and
 the solar neutrino solutions, 
 and $\alpha$ $(\sin \theta_{13})$ is the free 
 parameter. 
If $O( \alpha ) \simeq O( \beta )$, which 
 does not have physical meaning, 
 Eqs.(\ref{17}) induce
\begin{eqnarray}
\label{14p}
U_{e3} \simeq \frac{1}{4} \sin 2\theta_{12} 
      \left(
           \frac{\Delta m^2_{sol}}{\Delta m^2_{ATM}}
      \right) .
\end{eqnarray}
\vspace*{-0.2cm}
This equation is the same as that of 
 Ref.\cite{ABR}. 
As we have shown in the case of Type A, 
 Eq.(\ref{14p}) is the trivial equation 
 which is obtained from the   
 trivial assumption of $O( \alpha ) \simeq O( \beta )$, 
 since $\alpha$ is the free parameter which 
 has nothing to do with $\beta$.

\vspace{0.4cm}

Similar discussions can be applied to 
 the case of (b2), which is 
 $M_{\nu}^d \simeq diag.(m_1, -m_2, 0)$ in the first order. 
With the notation of 
 $m_2 = -m_1+d$ $(|d| \ll m_1)$,  
the neutrino mass matrix $M_\nu$ is given by 
\vspace{-0.2cm}
\begin{eqnarray}
M_\nu 
  &=&
  m_1
  \left(
  \begin{array}{ccc}
  (c^2-s^2)+\kappa & \ -\sqrt{2}cs+\alpha +\beta 
            & \sqrt{2}cs+\alpha -\beta \\
  -\sqrt{2}cs+\alpha +\beta 
      & -\frac{1}{2}(c^2-s^2) 
    +\delta +\gamma-\delta'  	
      & \frac{1}{2}(c^2-s^2)  +\delta-\gamma \\
  \sqrt{2}cs+\alpha -\beta
     & \frac{1}{2}(c^2-s^2)+\delta-\gamma
     & -\frac{1}{2}(c^2-s^2)
       +\delta +\gamma+\delta' 
  \end{array}
  \right) \nonumber \\
   \label{typeB-2}
\end{eqnarray}
where
$
\kappa \equiv{ds^2}/{m_1}$, 
$\alpha \equiv -({\epsilon}/{\sqrt{2}})\;[
 (c^2-s^2) - m_3/ m_1 + d s^2 / m_1]$, 
$\beta \equiv {dcs}/{\sqrt{2}m_1}$, 
$\delta \equiv{m_3}/{2m_1}$, 
$\gamma \equiv {dc^2}/{2m_1}$, 
$\delta' \equiv [{(d-2m_1)cs / m_1}] \epsilon$  . 

The values of $\kappa, \beta, \delta$, and $\gamma$ 
 are determined by the atmospheric and the solar
 neutrino solutions. 
The parameters $\alpha$ and $\delta'$ are
 free, which contain $\epsilon$. 
The mixing angles are induced from 
 Eq.(\ref{typeB-2}) as 
\begin{eqnarray}
\label{b22}
 \tan 2 \theta_{12}=-\frac{\sqrt{2}(\beta-\sqrt{2}cs)}
             {(c^2-s^2)-\gamma+\frac{\kappa}{2}}
      \;,\;\;\;\;\;
 \sin \theta_{13}=-\frac{\sqrt{2}\alpha}{(c^2-s^2)+\kappa-2\delta}.
\end{eqnarray}
By using the approximations of 
%
%\begin{eqnarray}
$
\Delta m^2_{ATM} \simeq  m_1^2 
     [( c^2-s^2-\gamma+{\kappa}/{2})^2 
     + 2 ( \sqrt{2}cs-\beta )^2 ] 
$ 
and 
$
\Delta m^2_{sol} 
   \simeq 4m_1^2(\gamma+{\kappa}/{2})
              [(c^2-s^2-\gamma+{\kappa}/{2})^2
               +2(\sqrt{2}cs-\beta)^2
             ]^{1/2}, 
$
the mixing angles in 
 Eqs.(\ref{b22}) become 
\begin{eqnarray}
& & \tan 2\theta_{12} \simeq -\sqrt{2}\beta, \;\;\;\;\;\;\;\;\;\;\;
    \sin \theta_{13} \simeq -\sqrt{2}\alpha ,   
               \;\;\;\;\;\;\;\;\;\;
                \;\;\;\;\;\;\;\;\;\;\;\;\;\;
               \;\;\;\;\;\;({\rm small \;\;mixing}), \\
& & \sin 2 \theta_{12} \simeq 1-\sqrt{2}\beta ,
      \;\;\;\;\;\;
 \sin \theta_{13} \simeq -{\sqrt{2}\alpha}
                           {\left(
                           \frac{d}{2m_1}-\frac{m_3}{m_1}
                           \right)^{-1}}, 
               \;\;({\rm large\;\;mixing}).
\label{b2} 
\end{eqnarray}
If $O( \alpha ) \simeq O( \beta )$, 
 Eqs.(17) and (\ref{b2}) induce 
\begin{eqnarray}
& & \sin \theta_{13} \simeq \tan \theta_{12}, 
  \;\;\;\;\;\;\;\;\;\;\;\;\;\;\;\;\;\;\;\;\;\;\;\;\;\;\;\;\;
  \;\;\;\;\;\;\;\;\;\;\;({\rm small\;\;mixing}),  \\
& & \sin \theta_{13} \simeq ({1-\sin 2 \theta_{12}})
            {\left(
              \frac{m_3}{m_1}-\frac{d}{2m_1}
             \right)^{-1}}, 
  \;\;\;\;\;({\rm large\;\;mixing}).
\end{eqnarray}
Similarly this is also the trivial relation.

\vspace{5mm}
% ======================================================================== %
%\section{Type C ($m_1 \sim m_2 \sim m_3$)}
% ------------------------------------------------------------------------ %

In Type C with the mass hierarchy of 
 $m_1 \sim m_2 \sim m_3$, 
 we show the case of (c4)\cite{HO1}, for example, 
 which is 
 $M_{\nu}^d \simeq diag.(m_1, m_2, m_3)$ in 
 the first order. 
With the notation of 
 $m_2=m_1+d$, $m_3=m_1+D$, and 
 $|d|\ll|D|\ll m_1$, 
 the neutrino mass matrix $M_\nu$ is given by 
\vspace{-0.3cm}
\begin{eqnarray}
M_\nu 
  &=&
  m_1
  \left(
  \begin{array}{ccc}
   1+\kappa 
       & \ \alpha +\beta 
            & \alpha -\beta \\
   \alpha +\beta & 1   +\delta +\gamma-\delta' 	& 
    \delta-\gamma \\
  \alpha -\beta  & \delta-\gamma & 1
      +\delta +\gamma+\delta' 
  \end{array}
  \right) \nonumber \\
    \label{typeC-4}
\end{eqnarray}
\vspace{-0.2cm}
where
$
\kappa \equiv{ds^2}/{m_1},\; 
\alpha \equiv ({\epsilon}/{\sqrt{2}})
       [
          {D}/{m_1}
          -{d s^2}/{m_1} 
        ], \;
\beta \equiv {dcs}/{\sqrt{2}m_1},\;
\delta \equiv {D}/{2m_1}, \;
\gamma \equiv {dc^2}/{2m_1}, \;
$ 
and
$
\delta' \equiv ({dcs}/{m_1}) \epsilon . 
$
Equation (\ref{typeC-4}) 
 induces the mixing angles as 
\begin{eqnarray}
\tan \theta_{12}=\frac{\sqrt{2}\beta}{\gamma-\frac{\kappa}{2}},
\;\;\;\;\;\;\;\;\;
\sin \theta_{13}=\frac{\sqrt{2}\alpha}{2\delta-\kappa}. 
\end{eqnarray}
\vspace{-0.1cm}
By using the approximations of 
%
%\begin{eqnarray}
$
\Delta m^2_{sol} 
     \simeq 4 m_1^2 (1+\gamma+{\kappa}/{2})
                [(\gamma-{\kappa}/{2})^2+2\beta^2]^{1/2}
$
and
$
\Delta m^2_{ATM} \simeq 4\delta m_1^2 ,
$
%\end{eqnarray}
%
we can obtain 
\begin{eqnarray}
\frac{\beta}{\delta}
   \simeq \frac{1}{\sqrt{2}} \sin 2\theta_{12}
      \left(
           \frac{\Delta m^2_{sol}}{\Delta m^2_{ATM}}
      \right), \;\;\;\;\;\;\;\;\;
\frac{\alpha}{\delta}
   \simeq \sqrt{2} \sin \theta_{13}. \label{typeC-ab}
\end{eqnarray}
Under the trivial assumption of 
 $O( \alpha ) \simeq O( \beta )$, 
 Eqs.(\ref{typeC-ab}) induce the trivial relation
\begin{eqnarray}
\sin \theta_{13} \simeq
      \frac{1}{2}\sin 2 \theta_{12}
        \frac{\Delta m^2_{sol}}{\Delta m^2_{ATM}}.
\end{eqnarray}

\vspace{5.0mm}
% ======================================================================== %
%\section{Summary}
% ------------------------------------------------------------------------ %

It has been said that the measurements of $U_{e3}$ in the 
 lepton flavor mixing matrix would help discriminate between 
 the possible solar neutrino solutions under the 
 natural conditions with the neutrino mass hierarchies of 
 $m_1 \ll m_2 \ll m_3$ and $m_1 \sim m_2 \gg m_3$ \cite{ABR}. 
However, it is not true, and 
 the relation between $\sin^2 2 \theta_{12}$ and $U_{e3}$ 
 obtained in Ref.\cite{ABR} is trivial in actual.

Why can not we obtain the relations 
 between the value of $U_{e3}$ and the solar neutrino solutions? 
This is easily understood as follows. 
Neglecting the $CP$ phases in the lepton sector, 
 the number of independent parameters in 
 the Majorana mass matrix of neutrino 
 are six. 
Five parameters are enough to determine the MNS
 matrix, since overall factor in the
 neutrino mass matrix does not contribute to 
 the MNS matrix. 
Thus we need five input parameters 
 in order to determine the MNS matrix, 
 and its element $U_{e3}$. 
Since the neutrino oscillation experiments except for
 the CHOOZ 
 give us only four input parameters 
 $\Delta m_{\rm ATM}^2$, $\Delta m_{\rm sol}^2$, 
 $\sin^2 2 \theta_{12}$, and $\sin^2 2 \theta_{23}$, 
 the value of $U_{e3}$ remains as 
 an unknown parameter, which we only know the upper bound 
 from CHOOZ experiments as
 $U_{e3}< 0.16$\cite{CHOOZ}. 
Therefore 
 the value of $U_{e3}$ cannot predict
 the solar neutrino solutions 
%In order to 
% predict the solar neutrino solution 
% from the value of $U_{e3}$,  
% the additional nontrivial experimental or 
% theoretical information must be needed. 
 without one additional nontrivial 
 condition.

\vspace{5mm}
\noindent
%\section*{Acknowledgment}
The work of NH is supported by the Grant-in-Aid for Science 
 Research, Ministry of Education, Science and Culture, Japan
 (No. 12740146, No. 12014208).

% ======================================================================== %
%
% Section bibliography
%
% ------------------------------------------------------------------------ %

\end{document}